\documentclass[conference]{IEEEtran}
\IEEEoverridecommandlockouts
\usepackage{cite}
\usepackage{amsmath,amssymb,amsfonts}
\usepackage{algorithmic}
\usepackage[ruled,vlined]{algorithm2e}
\usepackage{amsmath}
\usepackage{graphicx}
\usepackage{textcomp}
\usepackage{xcolor}
\def\BibTeX{{\rm B\kern-.05em{\sc i\kern-.025em b}\kern-.08em
    T\kern-.1667em\lower.7ex\hbox{E}\kern-.125emX}}
\begin{document}

\title{Causality Diagrams using Hybrid Vector Clocks}

\author{\IEEEauthorblockN{Ishaan Lagwankar}
\IEEEauthorblockA{Michigan State University, \\
East Lansing\\
MI}
\and
\IEEEauthorblockN{Kanishka Wijewardena}
\IEEEauthorblockA{Michigan State University, \\
East Lansing\\
MI}
}

\maketitle

\begin{abstract}
Causality in distributed systems is a concept that has long been explored and numerous approaches have been made to use causality as a way to trace distributed system execution. Traditional approaches usually used system profiling and newer approaches profiled clocks of systems to detect failures and construct timelines that caused those failures. Since the advent of logical clocks, these profiles have become more and more accurate with ways to characterize concurrency and distributions, with accurate diagrams for message passing. Vector clocks addressed the shortcomings of using traditional logical clocks, by storing information about other processes in the system as well. Hybrid vector clocks are a novel approach to this concept where clocks need not store all the process information. Rather, we store information of processes within an acceptable skew of the focused process. This gives us an efficient way of profiling with substantially reduced costs to the system. Building on this idea, we propose the idea of building causal traces using information generated from the hybrid vector clock. The hybrid vector clock would provide us with a strong sense of concurrency and distribution, and we theorize that all the information generated from the clock is sufficient to develop a causal trace for debugging. We post-process and parse the clocks generated from an execution trace to develop a swimlane on a web interface, that traces the points of failure of a distributed system. We also provide an API to reuse this concept for any generic distributed system framework.
\end{abstract}


\section{Introduction}

\begin{figure*}[h]
    \centering
    \includegraphics[width=\linewidth]{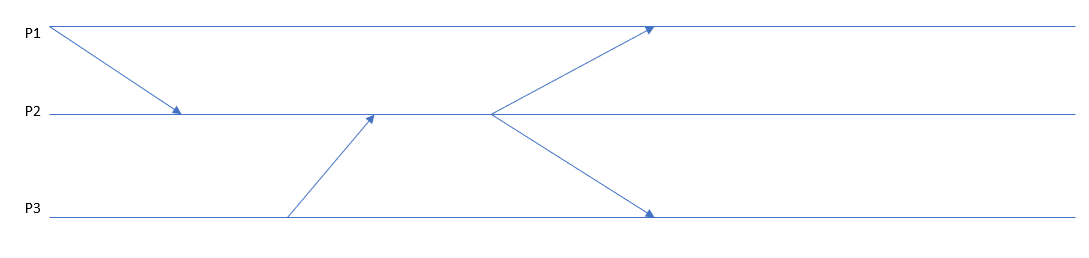}
    \caption{Example Swimlane Diagram with 3 Processes}
    \label{fig:swimlane_example}
\end{figure*}

Causality diagrams have long been used in distributed systems to produce relationships between how messages are communicated and interpreted by different processes in the system. They are often represented as swimlane diagrams, where message passing is represented as transactions between processes shown by horizontal axes, as depicted in figure \ref{fig:swimlane_example}.

Distributed systems often rely on clock systems to ensure consistency across replicas, debugging, and causal ordering of events to track execution. Clock systems in the new age have come a long way since their inception, and the strides taken in logical time to define causality in distributed systems. Vector clocks allow finer control over this causality, where each process independently tracks the clocks of the other processes in the system. This proves particularly useful when clocks exhibit \textit{skew} with respect to each other. The skew described is a characteristic of a system when the frequency of the quartz clock used to update the system clock changes over time, causing clocks to drift apart from each other. This was the main motivation of introducing logical time, but soon it was apparent that we would need physical time to better explain execution traces and attribute them to external factors due to real-world factors tampering with system execution. In the new age of networking, the Network Time Protocol (NTP) was a key factor in ensuring clocks remain within consistent bounds of each other, but this protocol proved to be a large memory bottleneck as the number of bits the clocks would use would be proportional to the number of processes in the system. This was a key factor in introducing clocks that reduced the number of bits the clocks needed to represent the causal consistency of a distributed system. One of the earliest works on this was the Hybrid Logical Clock, which allowed lesser bits to be stored while maintaining physical time on the system by rounding off high precision clocks to lower precisions, due to the properties of the distributed system that introduce latency. The latency induced allows us to ignore the higher precision offered by protocols such as NTP while preserving the happens-before relationship between messages.

We implement a version of the Hybrid Vector Clock \cite{b21}, a clock system that extends the idea of the HLC to vector clocks. In the HVC, we keep an array of offsets to keep track of potential skews in the distributed system. The key idea here is to store offsets rather than the large physical time, and for that, we discretize our time into intervals, based on a value set by the system administrator. For events between epochs, we implement a counter system to count events happening within an interval to make sure all events are tracked. 

Through the clock's logic, we can enforce causality if a clock of a process is lesser than a clock of another process. If two clocks are equal, those events are concurrent. If a process ${e}$ sends a message to another process ${f}$, the clock of ${e}$ is guaranteed to be lower than process ${f}$. This provides a way to enforce a partial order of the clock system. The concept of partial ordering has been explored widely in the concept of logical clocks, but through the HVC, we show that lesser information is sufficient to enforce the ordering. 

Using clocks to develop execution traces has been studied using older clock systems, such as Lamport clocks, vector clocks, and so on. These use the more primitive implementations of these clocks, where they store vast amounts of information to generate traces. This adds a high overhead on the distributed system, giving diminishing returns. A traditional vector clock, for example, stores data about all processes. This scales as a factor of the number of processes. We reduce this overhead to a subset of the number of processes, which are only the processes perceived to not be drifted away from the process point of view. We term this drift to be $\epsilon$, the clock skew. Using the HVC, we need only store how the clocks are offset in time from each other, and if the offset is too large, we need not need care about the time of that particular process. We can show that these clocks still enforce the same partial ordering, as any involvement from a skewed process would prompt the clock to record that interaction. Simply put, we care about skewed processes only when they take part in the distributed system's working. If we have not heard from a process, from the point of view of the clock, the process is considered dead, and we need not store information about it. This idea is particularly useful in an execution trace, as we can pinpoint processes not performing in a distributed computation, while also characterizing failure in those processes. This could also prove useful in determining what processes have skewed clocks in time-dependent computations.

In this work, we propose a system to generate execution traces and visualize them on an easy-to-use web interface. We provide controls to tweak how the traces are shown and characterized. We also provide an API to integrate the trace generation and visualization in the infrastructure it is implemented in. The rest of this paper is organized as follows. We first give a background on the advent of clocks and how distributed systems have changed using different clock infrastructures. We then provide the recent developments in generating execution traces from clock systems and motivate why our system would outperform the state of the art. We then provide our design of the clock and the proposed methodology to generate the execution traces.

\textbf{Our contributions: }We Implement the design of the Hybrid Vector Clock (HVC) proposed in \cite{b21} and expose it as an API. We then build a visualization infrastructure that utilizes the HVC infrastructure to show traces of execution with controls to tweak the visualization to observe process behavior during execution.

\section{Background}

\subsection{Clocks in Distributed Systems}

Logical clocks were proposed in 1978 by Leslie Lamport \cite{b1} to trace the ordering of events in a distributed system. Vector time was designed independently by multiple researchers \cite{b14}\cite{b2}\cite{b16}, and they proposed the idea of representing time in a distributed system as a set of n-dimensional non-negative integer vectors. According to \cite{b20}, Vector clocks are defined by three properties: Isomorphism, Strong consistency, and Event Counting. Isomorphism suggests that if two events $x$ and $y$ have timestamps $vh$ and $vk$, respectively, then $x \longrightarrow y \Longleftrightarrow vh < vk$. Here, $\longrightarrow$ implies a partial ordering between a set of events. Strong consistency implies that by examining the vector timestamp of two events, we can determine the causal relationship between the two events. Event counting suggests that if $d$ is always 1 in the rule $R1$, then the $i^{th}$ component of the vector clock at process $p_i, vt_i[i]$, denotes the number of events that have occurred at $p_i$ until that instant. 

There have been several prior implementations of vector clocks including Singhal-Kshemkalyani’s differential technique \cite{b17} and Fowler-Zwaenepoel direct dependency technique \cite{b18}. While vector clocks are upper bounded by $O(n)$ complexity in terms of both time and memory complexity, the different implementations of the past have tried to reduce this complexity and generate more efficient representations, with some success. Singham-Kshemkalyani's differential technique \cite{b17}, relied on piggybacking using the last sent and last update, without updating every vector clock. This method relies on the assumption that even though the number of processes is large, only a few key processes in a system would interact frequently by passing messages. A benefit of this method is that it cuts down storage overhead at each process to $O(n)$. However, this method doesn't make a substantial contribution to reducing the time complexity incurred when updating the vector clock, as it relies on piggybacking to work. 

Fowler-Zwaenepoel's direct-dependency technique cuts down storage complexity again by reducing the message size during transmission by transmitting only a scalar value in the messages. Here, a process only maintains information regarding direct dependencies on other processes. The downside of this method is that it has a high computational overhead as it has to trace dependencies and update the vector clock, especially in systems where a few key processes may have a large number of events. 

Clock synchronization using Network Time Protocol (NTP) uses the Offset Delay Estimation method to ensure physical clocks are synchronized across the internet. Clock offsets and delays are calculated, and timestamps are issued between different machines within a system accordingly. The system then attempts to establish a causal relationship by using the corrected timestamps. This however, can be computationally expensive and is open to error, as the delay estimation may not always be accurate and result in a violation of the causal relationship between processes issued by different machines. 

One existing limitation between vector clocks representing logical time and physical clock synchronization is the difficulty in reconciling one with the other. To overcome this challenge, Hybrid Logical clocks were introduced by Kulkarni et al. \cite{b13} to capture the causality relationship of a logical clock with the characteristics of a physical clock embedded into it. Another variant of the hybrid clock is the Hybrid Vector Clock \cite{b19}, \cite{b21}, which, unlike the Hybrid Logical Clock, is able to provide all possible/potential consistent snapshots for a given time. For this experiment, we use the Hybrid Vector Clock design presented in Yingchareonthawornchai et al. \cite{b21}, as it provides desirable characteristics to build our visualization framework.

\subsection{Visualizing Traces}

Mattern \cite{b2} talks about how distributed systems use the concept of global state to communicate information, and the need to characterize this global state. They talk about how a process can only approximate the global view of the system, and no process can have, at any given instant, a consistent view of the global state. To verify a distributed system, the author provides a comparison between three key approaches: simulating a synchronous distributed system given an asynchronous system, simulating a common clock or simulating a global state. They highlight the need of a vector clock system to provide a consistent snapshot of the global state, as each process having a clock that stores only its own state is not enough to describe the global state of the computation. 

PARAVER \cite{b3} uses the PVM message passing library to analyze traces generated from a computation. PVM primarily uses parallel message passing, and PARAVER analyses these parallel traces using data analytics and provides a graphical description of the analysis. This was one of the earliest visualization works on distributed systems simulated only for parallel traces. It used a parser to parse through the logs of the PVM-generated traces and analysed CPU activity, communications, and user events. This however required the addition of functionality to PVM itself and was not generalized to any distributed system interface. It also did not provide generic support and incurred a larger overhead to profile system resources while computation went on.

VAMPIR \cite{b4} provided analysis of MPI programs by generating timeline traces by profiling MPI applications. It used different visualization metrics to show whether processes were still active or not. It also provided views of system activities and aggregated statistics about the system itself. However, it was made specifically only for MPI applications and added to the profiling interfaces of MPI.

D3S \cite{b5} allowed developers to specify predicates on distributed properties of the system. These predicates can vary depending on what consistency checks one would require on the distributed system. They modeled the tracing as a consistency checker, and generated traces of predicate evaluation. The predicates are injected dynamically at compile time into the system and are evaluated based on the customization provided by the user. However, we believe this approach would add overhead to running the distributed computation due to complex predicate checking.

Zinsight \cite{b6} provides hierarchies of tasks and provides aggregated metrics to show timeline visualizations of events. It also provides users with changing granularity of the metric they want to see with sequences of computations per process. 

Trumper et al. \cite{b7} present a dynamic analysis tool that uses boundary tracing and post processing to analyse system behavior through a distributed computation. These are task based visualizations, where tasks are mapped to memory resources. But this may not always be the case where processes share the same memory, as in the case of OpenMP based infrastructures.

Dapper \cite{b8} is Google's tracing software for distributed systems where they provided low overhead, application-level transparency and scalability. Dapper uses annotations and \textit{spans} to generate traces through RPCs. However, the authors mention that Dapper cannot correctly point to causal history, as it uses annotations in non-standard control primitives, that may mislead the causality calculations. Our approach would overcome this, as causality is enforced through a lattice of clocks, rather than the events itself.

Isaacs et al. \cite{b9} provide a comprehensive survey of different distributed monitoring and tracing tools in the past decade, providing detailed descriptions and categorizations based on task parallelism, causality information and so on.

Isaacs, Bremer et al. \cite{b10} design a trace visualization system purely relying on logical clocks, and then transposing those clocks back to real-time clocks in the visualization. Processes are also clustered based on logical behavior. However, this would incur more overhead than our solution, and may cause conflicts in enforcing causality, due to the usage of a standard logical clock.

Verdi \cite{b11} provides developers with choosing the fault system to diagnose, and verifies the implementation of the system. This is a formal verification system where it provides the developer with an idealized fault model, and once this is verified, it applies the correctness to a more realistic fault model.

ShiViz \cite{b12} uses vector clocks in generating distributed system traces using happens-before relationships. By using vector clocks, it provides a verifiable and accurate notion of causality. However, since it uses traditional vector clocks, it uses a higher complexity than our proposed model. 

\section{Hybrid Vector Clocks}

In this section, we provide the design of the clock we would be using for generating causal traces. The hybrid vector clock (HVC) \cite{b21} stores \textit{offsets} instead of the timestamp themselves. This is very useful, as instead of storing logical time, we can store the physical time of a particular process and its offset with another process, providing the physical time of all processes in the system. We ignore skewed processes that have drifted too far away from the process' point of view, using $\epsilon$, the clock drift. 

 \begin{figure}
     \centering
     \includegraphics[width=\linewidth]{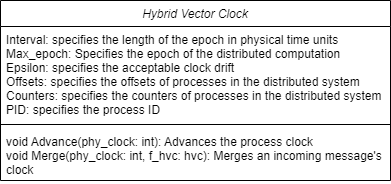}
     \caption{HVC Class Diagram}
     \label{fig:hvc_class_diagram}
\end{figure}

Figure \ref{fig:hvc_class_diagram} shows the class diagram of the hybrid vector clock. The \textit{offsets} store the offset of all other clocks in the point of view of the process clock. The \textit{counters} store the counters of events occurring in the same epoch. 

We enforce partial ordering of all processes in the system, such that they form a lattice of clocks. The way we enforce the order is as follows:

Assume for a HVC of process $i$, there exists an array clock$_{hvc}$: \\
\begin{equation}
    max\_epoch + offset[i] + counter[i] = clock_{hvc}[i]
\end{equation}

Then, for two events ${e}$ and ${f}$, we have the following relation, where $\rightarrow$ implies causality between ${e}$ and ${f}$:

\begin{gather*}
        e \rightarrow f \leftrightarrow \\
        \forall\ i\ in\ e.clock_{hvc}, 
        e.clock_{hvc}[i] \leq f.clock_{hvc}[i] \\
        \exists\ k\ in\ e.clock_{hvc}, 
        e.clock_{hvc}[k] < f.clock_{hvc}[k]       
\end{gather*}

We have two key algorithms in the clock's execution. The \textbf{Send (Or Local event)} [Algorithm \ref{alg:send}] algorithm specifies how a clock moves forward with respect to the system clock. The \textbf{Recv} [Algorithm \ref{alg:receive}] specifies how an incoming clock from a message updates the process' clock. This allows each clock to independently update itself, rather than receiving inputs from the infrastructure. This provides a system with no deadlocks and allows asynchronous execution.

\begin{algorithm}
\KwData{\\
e: hybrid clock of event e
phy\_clock: physical clock of the process
}
\KwResult{f: The next hybrid clock after one step}

$new\_max\_epoch \leftarrow max(e.max\_epoch, floor(phy\_clock/e.interval))$\\
\eIf{$new\_max\_epoch == e.max\_epoch$}{
    $f.counters \leftarrow e.counters$\\
    $f.counters[f.pid] += 1$\\
}{
    $f.counters \leftarrow [0]*N$\\
    \For{$idx \leftarrow 0$ \KwTo  $N$}{
     // To be converted to O(number of non-nulls) algorithm\\
    $f.offsets[idx] \leftarrow min(e.offsets[idx] + new\_max\_epoch - e.max_epoch, \epsilon)$
    }
}

$f.max\_epoch = new\_max\_epoch$
$f.offsets[f.pid] = 0$

\Return{f}

\caption{Send Event}
\label{alg:send}
\end{algorithm}

\begin{algorithm}
\KwData{\\
m: hybrid clock of incoming message event m\\
e: hybrid clock of event f\\
phy\_clock: physical clock of the process\\
}
\KwResult{f: The next hybrid clock after one step of event f}

$new\_max\_epoch \leftarrow max(e.max\_epoch, m.max\_epoch, phy\_clock/e.interval)$\\
\eIf{$new\_max\_epoch == e.max\_epoch\ and\ new\_max\_epoch == m.max\_epoch$}{
    \For{$pid \leftarrow 0$ \KwTo  $N$}{
     // To be converted to O(number of non-nulls) algorithm\\
        $f.counters[pid] \leftarrow max(e.counters[f.pid], m.counters[pid])$\\
    }
    $f.counters[f.pid]\ =\ max(e.counters[f.pid], m.counters[f.pid])\ +\ 1$\\
    $f.max\_epoch = new\_max\_epoch$\\
    $f.offsets[f.pid] = 0$
}{
    \eIf{$new\_max\_epoch == e.max\_epoch$}{
        $f.counters[f.pid] += 1$\\
        $f.offsets[m.pid] = floor(min(e.max\_epoch - m.max\_epoch, \epsilon))$
    }{
        \eIf{$new\_max\_epoch == m.max\_epoch$}{
            $f.offsets = m.offsets$
            $f.counters = m.counters$
            $f.counters[f.pid] += 1$
            $f.max\_epoch = m.max\_epoch$
        }{
            $f.advance(phy\_clock)$
        }
    }
}

$f.offsets[f.pid] = 0$

\Return{$f$}

\caption{Recv Event}
\label{alg:receive}
\end{algorithm}

Here, we illustrate the working of the HVC through some sample scenarios. 

\begin{figure}
    \centering
    \includegraphics[width=\linewidth]{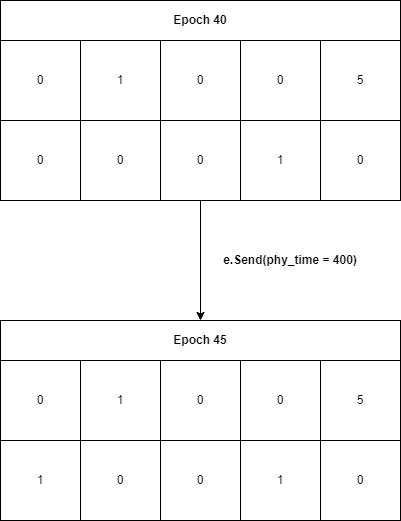}
    \caption{Advancing when the epoch does not change.}
    \label{fig:advance-1}
\end{figure}

\begin{figure}
    \centering
    \includegraphics[width=\linewidth]{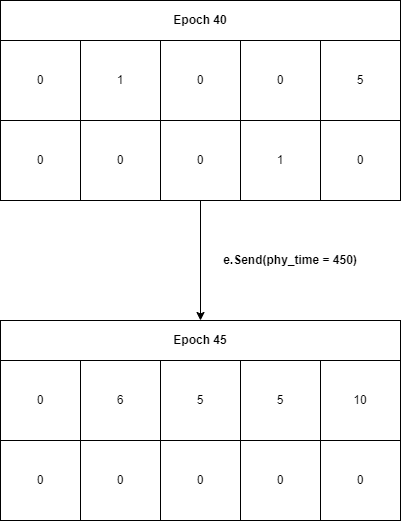}
    \caption{Advancing when the epoch changes.}
    \label{fig:advance-2}
\end{figure}

\begin{figure}
    \centering
    \includegraphics[width=\linewidth]{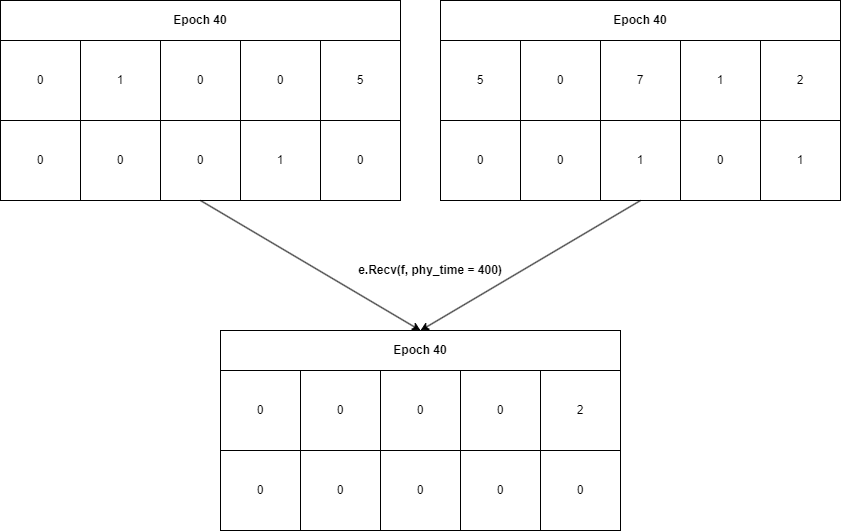}
    \caption{Merging when both the message and process are in the same epoch.}
    \label{fig:merge-1}
\end{figure}

\begin{figure}
    \centering
    \includegraphics[width=\linewidth]{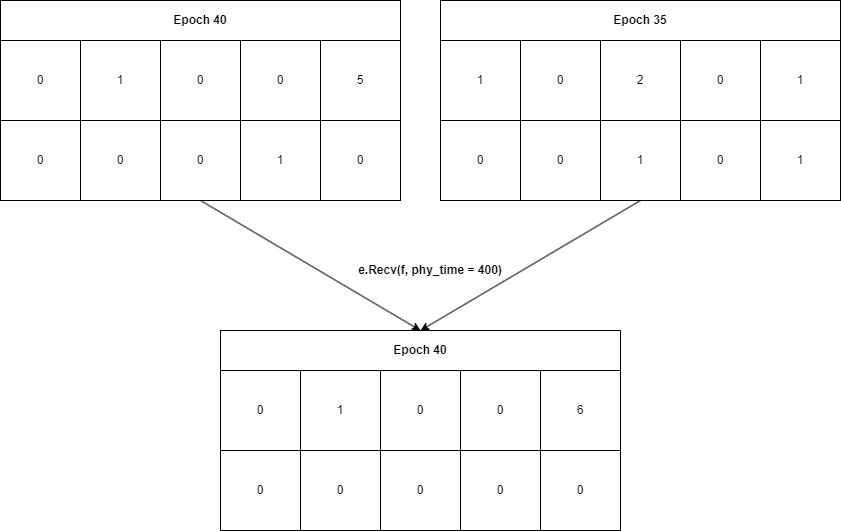}
    \caption{Merging when the message is lagging.}
    \label{fig:merge-2}
\end{figure}

\begin{figure}
    \centering
    \includegraphics[width=\linewidth]{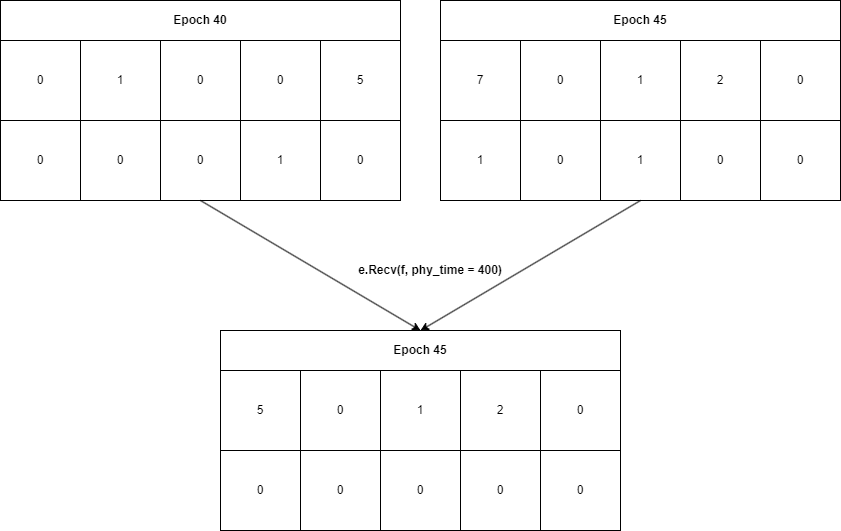}
    \caption{Merging when the message is ahead of the process.}
    \label{fig:merge-3}
\end{figure}

\begin{itemize}
    \item[1.1] \textbf{[Send] Physical clock of process in the same epoch}: In figure \ref{fig:advance-1}, the epoch the system is in does not change. In this case, we increment the counter of the process clock to show that an event has occurred.
    
    \item[1.2] \textbf{[Send] Physical clock of process in a different epoch}: In figure \ref{fig:advance-2}, the epoch the system is in changes. In this case, we update the max epoch of the clock. We do not make any other changes as with respect to the process, the other processes are leading/lagging by the same amount.
    
    \item[2.1] \textbf{[Recv] Message and process are in the same epoch}: In figure \ref{fig:merge-1}, the epoch of both the message and process are in the same epoch. In this case, we take the max of all pairwise counters in the message and process HVCs. We then increment the counter of the process in the clock.
    
    \item[2.2] \textbf{[Recv] Message is lagging with respect to the process}: In figure \ref{fig:merge-2}, the epoch of the message is lagging with respect to the process. In this case, we increment the counter of the process clock, and set the offset of the process that the message came from to the difference of the max epochs, or epsilon, depending on which is smaller.
    
    \item[2.3] \textbf{[Recv] Message is leading with respect to the process}: In figure \ref{fig:merge-3} the message's max epoch is much higher than that of the process. We copy over the offsets and counters to the process clock. Then we increment the counter, and then set the offset of the process to 0 to indicate that the process has caught up with the distributed system.

\end{itemize}

In all cases, we can observe that partial ordering is maintained. If any clocks come up with the same offsets, counters and max epoch, these events are deemed to be concurrent.

\section{Design and Implementation}

\subsection{Experimental Design}

To create our Hybrid Vector Clock implementation we utilized a Python 3.10 test-bed with each HVC being called from a custom-built class in Python. Our HVC class consisted of the following member variables:
\begin{enumerate}
    \item Interval: Integer (Representing the Epoch Interval)
    \item PID: Integer (Representing the Process ID)
    \item Epsilon: Integer (Representing Epsilon of the System)
    \item Max Epoch: Integer (Representing Epoch of the Process)
    \item Counters: HashMap (Representing Counters of the clock)
    \item Offsets: HashMap (Representing offsets of the clock)
\end{enumerate}

Each clock is initialized with the Interval, PID and Epsilon input by the user, with Max Epoch set to 0 and Counters and offsets set to empty. The reason for this is that the epoch, counters and offsets are updated based on the advance and merge events the processes are subjected to, and should not be hard-coded into the clocks to represent the accurate output when enforcing causality. We implemented unit-testing to verify that our outputs are accurate, where the max epoch, counters and offsets of the output clock of each merge and advance event was compared with the target clock to verify the accuracy. This allowed us to ascertain the accuracy of our clock class implementation to enforce causality from our input data. 


\subsection{Ordering Input Traces}

We used the idea of the HVC to embed information about generating traces. We introduced a new variable \textit{broken}, that would indicate a failure in sending a message or a system-level failure. This turned out to be the \textit{explicit} failure we would be graphing. \textit{Implicit} failures on the other hand, such as node power being lost mid computation would be tracked by polls. If a process stops responding to polls, we deem that process lost and set the process's broken bit to true. Using this information, we built a parser that parses logs and generates causal traces in physical time. This provides much better causal tracing as logical time would not make sense to a system administrator. With physical timestamps and partial ordering, we did a time-based analysis of system outages, and attribute them to real-world events that may have potentially impacted the system.

The process of sorting the time traces of the input data adhered to the algorithm seen in Algorithm \ref{alg:sort_traces}. The web interface was designed using the chart.js library and a sample preview of the web interface can be seen in Figure \ref{fig:api_example}. 

\begin{algorithm}
\KwData{\\
from\_node: Process ID of Message Transmitter\\
to\_node: Process ID of Message Receiver \\
sender\_max\_epoch: Epoch of Transmitter \\
sender\_offsets: Offsets of Transmitter \\
sender\_counters: Counters of Transmitter \\
output\_max\_epoch: Max epoch of output clock \\
output\_offsets: Offsets of output clock \\
output\_counters: Counters of output clock \\
output\_epsilon: Epsilon of system \\
epoch\_interval: Epoch Interval of system \\
}
\KwResult{output: An ordering of all the clocks in a causal format}
Read Input Data Input traces\\
Find send clock and output clock of input traces according to the following formulae:
$send\_time[i] = sender\_max\_epoch + sender\_offsets[i]  + sender\_counters[i]$\\
$output\_time[i] = output\_max\_epoch + output\_offsets[i]  + output\_counters[i]$\\
Order traces based on send time and output time according to the following formula: \\
if $send\_time$ $a < send\_time$ $b$ $\mathbf{then}$ \\
a appears before b \\
if $send\_time$ $a == send\_time$ $b$ $\mathbf{then}$ \\
\hspace*{5mm} if $output\_time$ $a < output\_time$ $b$ $\mathbf{then}$ \\
\hspace*{10mm} a appears before b \\
output: list containing ordering \\
\Return{output}

\caption{Sorting Input Traces}
\label{alg:sort_traces}
\end{algorithm}

\subsection{Designing the Web interface}

We designed a web interface that reads from the logs generated on the master node. We assume for the sake of this project that all clocks write to the same log base but provide a methodology that extends this approach to multiple systems. The abstraction would read from the single source database, and then construct timelines for each participating process. We designed the web interface using the json library in the Python programming language. The API has the following features and benefits: 

\begin{figure*}[h]
    \centering
    \includegraphics[width=\linewidth]{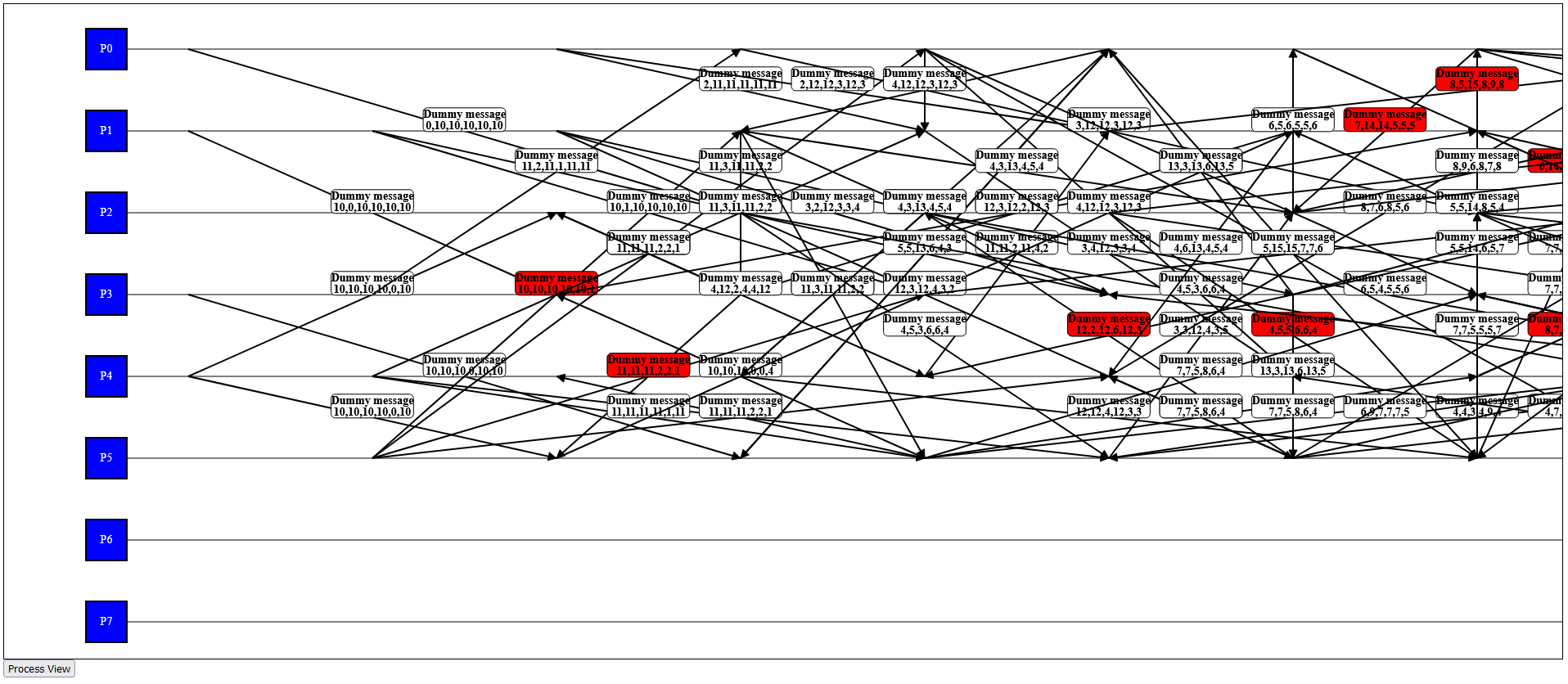}
    \caption{Sample API example of output traces with swimlanes in visualization}
    \label{fig:api_example}
\end{figure*}

\begin{enumerate}

\item The API was capable of isolating a particular message and zooming in on a visualization of a snapshot that isolated a subset of messages from a trace. This subset included some success cases but also highlighted failure cases and the reasons behind said failure. As can be seen in Figure \ref{fig:example_failure_api}, we can see some instances of a failure cases that was isolated from the main figure. 

\begin{figure*}[h]
    \centering
    \includegraphics[width=\linewidth]{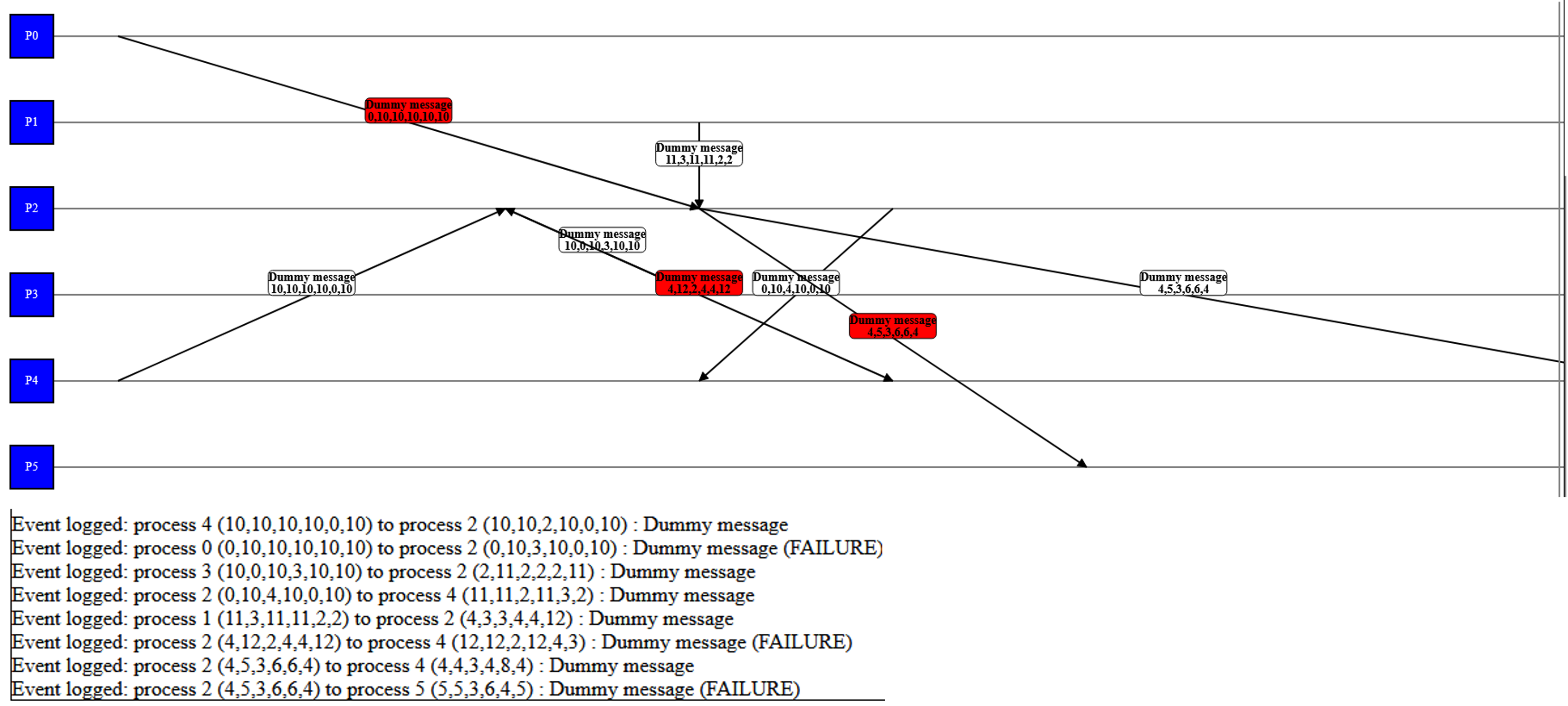}
    \caption{Sample API example of failure diagnosing in a single trace}
    \label{fig:example_failure_api}
\end{figure*}

\item The API is also dynamic in that it is capable of updating in real time. The API is connected to the localhost of the user machine, so it can be generated on an available port of the machine and then updated in real-time, based on updates made to the input data traces. 

\end{enumerate}

\section{Results and Discussion}

\subsection{Failure Diagnosing}

We evaluated the degree to which our system was capable of diagnosing failure and understanding causal violations in logic. As can be seen in Figure \ref{fig:example_failure_api}, this is a sample instance of failure that was observed in Process 2. As can be seen in each of the ``Failure'' messages, a dummy instance of failure was logged in for three out of the eight messages originating to and from Process 2. By observing the ``dummy'' messages, we could analyze the root cause of failure to see where a violation in causal consistency took place. 

\subsection{Memory and Time Complexity}

\begin{figure*}[h]
    \centering
    \includegraphics[width=\linewidth]{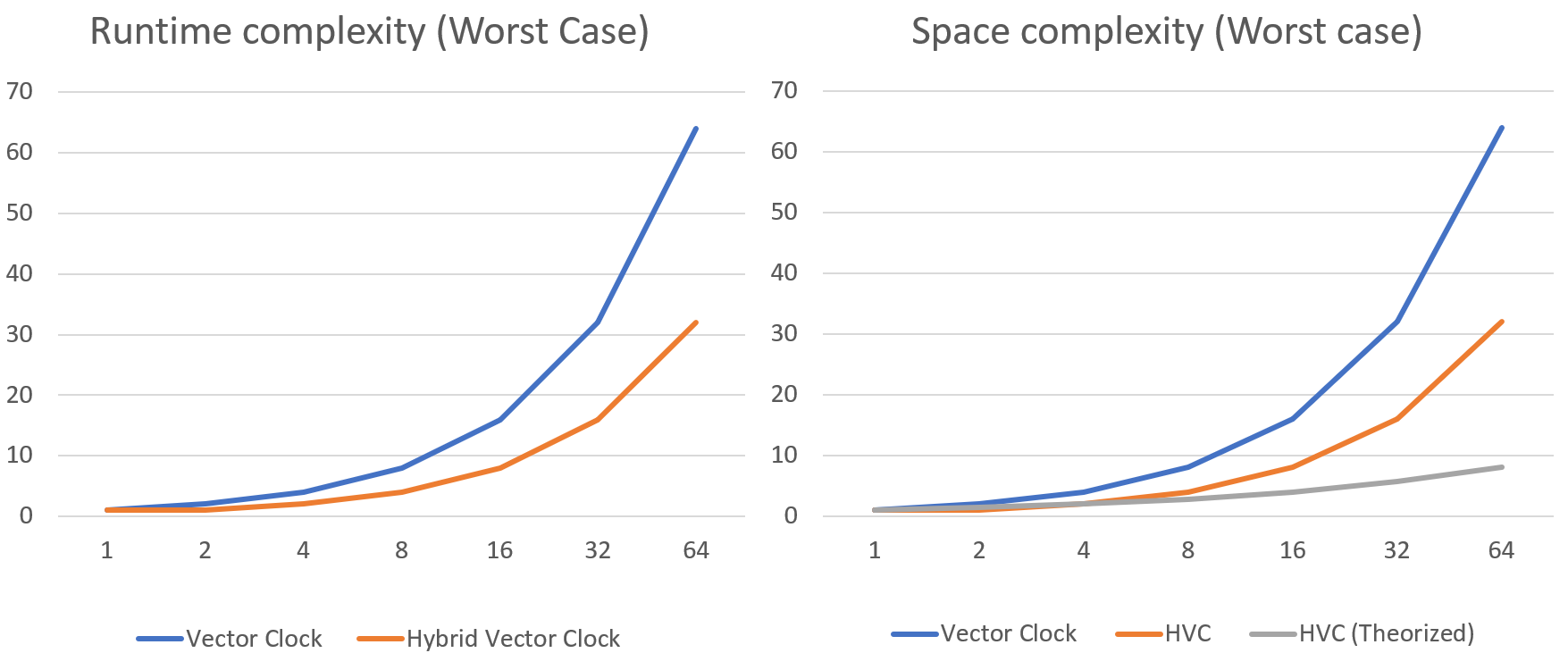}
    \caption{Runtime Complexity and Space Complexity of our HVC implementation compared to regular vector clock implementations}
    \label{fig:complexity}
\end{figure*}

We plotted the time complexity of our HVC implementation against traditional vector clock implementations and discovered that our HVC implementation, despite being bounded by $O(n)$ for the worst case scenario, generally outperformed a typical vector clock by a wide margin. In practical applications, our HVC had a performance close to $O(n/2)$, which is a much more efficient implementation than a Vector Clock with O(n) performance. We also found that the space complexity of our HVC implementation was bounded by $O(n/2)$. However, according to Kulkarni et al.\cite{b19}, the memory complexity of the HVC should be bounded by $O(\sqrt{n})$. This implies that there is more work to be done, especially in representing zero counters or offsets as empty in our hashmap representations. Therefore, hashmaps with empty counters and offsets should automatically be deleted from the HVC, which does cause issues during implementation as the counters and offsets would have to be re-instantiated each time a non-zero value is obtained for a counter or an offset, which means that it would be more convenient to simply represent the counters and offsets as hashmaps with no existing values, and accept the space complexity that comes with being marginally greater than the ideal space complexity.  

\subsection{Comparison against othe visualization schemes}

While our visualization scheme is not as robust or detailed as ShiViz \cite{b12} or D3S \cite{b5}, our ability to diagnose failure by isolating a single trace and our ability to visualize causality using swimlanes does indicate great promise in our implementation. We are optimistic that in future work we would be able to expand upon our API and refine it further, for instance to identify the full history of a process and diagnose where a point of failure took place. 

\section{Conclusion and Future Work}

We designed a visualization scheme for Hybrid Vector Clocks. We illustrated why HVCs were an improvement in the state-of-the-art when representing logical time in distributed systems, and then formulated a scheme that sorted input traces and then visualized the clocks built out of these input traces to see if our input traces enforced causality. Our visualization API was able to isolate traces with points of failure and we were able to evaluate causes of failure within a trace based on its history. 

Future work done in this regard would involve expanding upon the API to make it more robust. Improvements include being able to visualize a more in-depth history of a trace, and a visually appealing API compared to the state-of-the-art.






\end{document}